\documentstyle[hasson]{article}
\begin{document}
\title{
\vfill
A Bayesian test for the appropriateness of a model in the
biomagnetic inverse problem}

\author{R. Hasson\thanks{Applied Mathematics department}\ \ and 
S.J. Swithenby\thanks{Physics department}\ ,\\
The Open University,\\
Milton Keynes, \\
MK7 6AA\\
UK}
\maketitle

\vfill

\begin{abstract}
This paper extends the work of Clarke \cite{e:Clarke94} on the Bayesian
foundations of the biomagnetic inverse problem. It derives expressions for the
expectation and variance of the {\em a posteriori} source current probability
distribution given a prior source current probability distribution, a source
space weight function and a data set. The calculation of the variance enables
the construction of a Bayesian test for the appropriateness of any source
model that is chosen as the {\em a priori} infomation. The test is illustrated
using both simulated (multi-dipole) data and the results of a study of early
latency processing of images of human faces. 

\vfill

\subsubsection*{Keywords}
Biomagnetic inverse problem, Bayesian.
\end{abstract}

\vfill
\newpage

\section{Introduction}
The magnetoencephalographic (MEG) inverse problem (sometimes known as the
biomagnetic inverse problem) is the process of identifying the source current
distribution inside the brain that gives rise to a given set of magnetic field
measurements outside the head. The problem is difficult because the detectors
are limited in number and are sensitive to widespread source currents, and
because of the existence of silent and near-silent sources. 'Silent sources'
are configurations of current density inside the brain which give zero
magnetic field outside the head (e.g. all radial sources are silent when the
head is modelled as a conducting sphere). It follows that the general
biomagnetic inverse problem is ill-posed and under-determined. 

The most common way of reducing the problem and rendering it tractable is by
characterising the source in terms of a limited number of effective current
dipoles. Such source descriptions provide links with the
dominant functional architecture model of the brain in which processing is
described in terms of localised activity with interactions between essentially
separate regions. Multiple dipole models have enjoyed considerable success in
the analysis of sensory and motor cortex (e.g. 
\cite{e:Vanni96,e:Buchner95,e:Mauguiere97,e:Hoshiyama97}).

Growing evidence for the existence of more diffuse brain networks have led to an interest in distributed source algorithms. Several have been proposed 
\cite{Hamalainen84,e:Hamalainen94,p:Ioannides89,e:Wang92,e:Marqui94,e:Gorodnitsky95}. These algorithms have been designed to cope with the non-uniqueness 
of the problem, primarily by restriction of the source space and by
regularisation. Each algorithm leads to a unique solution (from the infinite
number available) through its particular choice of source basis, weight
functions, noise assumptions, and, in many cases, cost function. There has
been an extended debate about the accuracy and value of these methods. This
proceeds at two levels; the technical ability of the various algorithms to
recover a simulated source distribution (often quoted in terms of one or more
source current dipoles), and the electrophysiological appropriateness of the
type of source structure favoured by particular algorithm parameters. So, for
example the simple minimum norm solution \cite{e:Hamalainen94}, which tends to produce a grossly
smeared and superficial source distribution may be compared with the LORETA
solution \cite{e:Marqui94} which favours smooth but regionally confined current distributions.
The issues have been fully debated in recent conferences 
\cite{e:ISBET-loreta,e:Wood98}).

The many to one nature of the mapping of sources to magnetic fields suggests
that a probabilistic approach to reconstructing the sources from the magnetic
field could be used. A Bayesian probabilistic approach was first proposed by
Clarke \cite{e:Clarke94}. More recently, Baillet et al have described an
iterative approach which combines both spatial and temporal constraints within
a unified Bayesian framework designed to allow the estimation of source
activity that varies rapidly with position, e.g. across a sulcus
\cite{e:Baillet97}. Schmidt et al have developed a probabilistic algorithm in
which a bridge is made between distributed and local source solutions through
the use of a regional descriptor within the source representation 
\cite{e:Schmidt97}. In this case, a Monte Carlo method is used in the
absence of an analytic solution for the expectation value of the source
current. 

Here, we are proposing an alternative Bayesian approach. It includes the explicit assumption of both a prior source current 
probability distribution and a source space weight function, and allows the 
calculation of the expectation and variance of the a-posteriori source current 
probability distribution. The derivation of these quantities is detailed in Section 3.
The inclusion of the prior probability and the calculation of the variance provide a means by which the consistency of a model (assumed as the prior) can be tested against the data to reveal those parts of the source distribution that are statistically robust and, conversely, where the model is inadequate. Numerical calculation of significance is possible. A straightforward extension of this idea is the direct comparison of two data sets to reveal where, within a given model, there are significant 
differences in their associated source distributions.
In the final part of this paper, both simulated and real data will be used to illustrate these various uses of the technique.

\section{Specification of the problem}

\begin{figure}[htbp]
\begin{center} 
\mbox{\epsffile{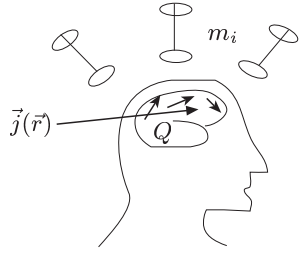}}
\end{center} 
\caption{\label{fig:problem} A schematic experimental geometry.}
\end{figure}

The arrangement of sources and detectors for the biomagnetic inverse
problem is shown in Figure~\ref{fig:problem}. The sources giving rise to the 
measurements are assumed to be restricted within a source space $Q$, which 
may be smaller than the whole head volume (e.g. if the sources are assumed to 
be cortical). The current density within the volume $Q$ is assumed to belong 
to the space of square integrable vector fields on $Q$, which we call $J$. 

The measurement process typically gives successive sets of data ($\sim 100$
channels) every millisecond. In this paper the data for each time instant is
processed independently, and the data from a single time instant is
collected into a vector $\m \in \R^N$. If $\jr \in J$ then the measurement
process can be represented by a functional $z:J\rightarrow \R^N$. 
A subscript notation will be used to identify the sensor, i.e. $z_i$ is the
ideal reading from the $i$th sensor. So the basic equation is: 
\[ \m_i = z_i (\j ) +e_i \] where the $e_i$ are the measurement errors, which 
are assumed to be normally distributed with zero mean and covariance matrix 
$\alpha^2 D$ (where $\alpha$ is the standard deviation of the errors and $D$ 
is a symmetric, positive definite matrix).

To compute the functional $z(\ )$ on a computer (i.e. to solve the forward
problem) requires a volume conductor model of the head. In this paper the
precise model used is irrelevant, so the final results will be written in
terms of $z$. This is done is via the Green's functions \Li\ for the problem,
which are defined by 
\[
z_i (\j) = \int_Q \Li \ofr \cdot \jr \dr
\]

Stated simply, the inverse problem is to estimate \jr\ given the data vector
$\m$. Obviously the given data are not enough to determine \jr\ uniquely (for 
several reasons). The approach adopted here starts from the same point as used 
in Clarke \cite{e:Clarke94}, a statement of Bayes's theorem:
\[
\label{eqn:orignialbayes}
\P (\j \in A | m \in B) = 
\frac
  {
  \P(m \in B | \j \in A) \P(\j \in A)
  }
  {
  \P (m \in B)
  }
\]
where $A$ is a set of currents and $B$ is a set of measurements. This equation
reads, the {\it a posteriori} probability of a current set $A$ after the
measurement $B$ is proportional to the probability of producing the
measurement $B$ given that the current is in the set $A$ times the {\it a
priori} probability of the current set $A$. ($\P (m \in B)$ is a constant for
any measurement set $B$). 

In this paper the probability distributions (both the prior probability and
the errors) are assumed to be gaussian and so it is permissible to work with
probability density functions. A further simplification is achieved by
shrinking the measurement set $B$ to a single point $\{ \m \}$ (this ignores
the finiteness of the precision of the measurements). Equation 
\ref{eqn:orignialbayes} then becomes: 
\[
\label{eqn:bayes}
\rho_{\m} (\j) \propto \rho (\j) \; \times \; \epsilon (\m - z(\j)) 
\]
where $\rho$ is the {\it a priori} distribution, $\rho_{\m}$ is the
{\it a posteriori} distribution and $\epsilon$ is the error distribution. 
Throughout the paper,
probability density functions will only be determined up to a constant. The
constant of proportionality is found by requiring that the probability is
normalised to one. In 
this paper both $\rho$ and $\epsilon$ are assumed to be Gaussian and then 
$\rho_{\m}$ is calculated to be Gaussian. An error probability density function 
$\epsilon$ consistent with the Gaussian assumption may be written:
\[
\label{eqn:error}
\epsilon (e) \propto
\exp \left\{ -\frac1{2\alpha^2} \trn{e} \inv{D} e \right\}
\]

This generally valid expression will be retained throughout the derivation in 
this paper. In practice, the noise covariance matrix $D$ may be difficult to 
estimate and, for simplicity, the simple form $D=I$ will be used in the
later illustrations.

\section{Derivation}

In this section formulae for the maximum likelihood current distribution and
also the expected error distribution are derived under specific assumptions.
First, an inner product on $J$ is defined by:
\[
\ip{\j_1}{\j_2} = \int_Q \frac{\j_1 \ofr \cdot \j_2 \ofr}{\omega \ofr} \dr
\]
where $\omega \ofr$ is a weighting distribution defined on the source space 
$Q$. This provides a method of inputting prior information of the location of 
sources (e.g. gained from MRI images) into the algorithm.

Clarke~\cite{e:Clarke94} assumed that the maximum likelihood prior current
density was identically zero. Here that restriction is avoided and an
arbitrary prior current \jp\ will be introduced as a parameter of the
method. The {\it a priori} probability distribution on $J$ is defined
using \jp\ and the inner product:
\[
\label{eqn:apriori}
\rho (\j) \propto \exp 
  \left\{ -\frac1{2\beta^2}  \ip{\j-\jp}{\j-\jp} \right\}
\]
where $\beta$ is the assumed {\it a priori} standard deviation.

To proceed further a basis is needed for $J$. A `natural' choice is a basis 
that is related to the measurement functional $z$. So an obvious candidate
is a basis derived from the adjoint map to the measurement map from $J$ to
$\R^N$. This gives a map $\V{z}:\R^N\rightarrow J$ (since $J$ is self-dual)
defined by:
\[
\ip{\V{z}(a)}{\j} = \trn{a}z(\j)
\]

Explicitly this gives the set of linearly independent distributions $\{ \omega
\ofr \Li \ofr \}$. This set is extended into a basis of
$J$ that includes the silent sources by adding vectors $\{ \li \}$ which are 
chosen to be orthogonal to the $\{ \omega \Li \}$ i.e. 
\[
\ip{\omega \Li}{\lj} = 0 \hspace{1cm} \forall i,j
\]
Since $\{ \omega \ofr \Li \}\bigcup\{ \li \}$ is a basis of $J$ a 
general current density \jr\ can be written in terms of this basis as
\[
\jr = \sum_{i=1}^N a_i \omega \ofr \Li \ofr + \sum_i b_i \li \ofr
\]
To simplify the notation the components of currents are written
in column vector notation:
\[
\j = \col{a}{b},\qquad\jp=\col{\prior{a}}{\prior{b}}
\]
A simple computation gives 
\[
\ip{\j - \jp}{\j - \jp} = 
  \trn{\col{a-\prior{a}}{b-\prior{b}}}
  \diag{P}{Q}
  \col{a-\prior{a}}{b-\prior{b}}
\]
where $P_{ij}=\ip{\omega \Li}{\omega \Lj}$ and 
$Q_{ij}=\ip{\li}{\lj}$ since by construction $\ip{\omega \Li}{\lj}=0$.

Now the two {\it a priori} probability density functions 
(Equation~\ref{eqn:apriori} and Equation~\ref{eqn:error}) may be combined 
with Bayes's theorem (Equation~\ref{eqn:bayes}) to obtain the 
{\it a posteriori} probability density: 
\[
\rho_{\m} (\j) \propto
\exp 
  \left\{
    -\frac1{2\beta^2} 
    \trn{\col{a-\prior{a}}{b-\prior{b}}}
    \diag{P}{Q}
    \col{a-\prior{a}}{b-\prior{b}}
  \right\}
\exp \left\{ -\frac1{2\alpha^2} \trn{(\m-z(\j))} \inv{D} (\m-z(\j)) \right\}
\]

The task now is to manipulate this equation so that it is explicitly in the
form of a gaussian distribution. As a first step the exponentials are combined:
\[
\rho_{\m} (\j) \propto \exp \left\{ 
    -\frac1{2\alpha^2} \left[
    \trn{\col{a-\prior{a}}{b-\prior{b}}} \diag{\zeta P}{\zeta Q} 
    \col{a-\prior{a}}{b-\prior{b}}
    +\trn{(\m-Pa)} \inv{D} (\m-Pa) \right]
  \right\}
\]
where $\zeta = \alpha^2/\beta^2$ and $z(\j)$ has been replaced by $Pa$. Next, 
the terms involving operators on $a$ are simplified by completing the square 
All constant terms can be absorbed into the normalization constant of the 
probability density function and are ignored in this derivation.
\<
\mbox{Operators on $a$}
& = \trn{(a-\prior{a})}\zeta P(a-\prior{a})+\trn{(\m-Pa)} \inv{D} (\m-Pa) \\
& = \trn{a}\zeta Pa-2\trn{a}\zeta P\prior{a}
  +\trn{a}P\inv{D}Pa-2\trn{a}P\inv{D}\m+\mbox{const.}\\
& = \trn{a}\left( \zeta P+P\inv{D}P\right)a
  -2\trn{a}[P\inv{D}\m+\zeta P\prior{a}]+\mbox{const.}\\
& = \trn{(a-\tilde{a})}\left(\zeta P+P\inv{D}P\right)(a-
\tilde{a})+\mbox{const.}
\>
where $\tilde{a}$ is defined by:
\<
\left(\zeta P+P\inv{D}P\right)\tilde{a}&=P\inv{D}\m + \zeta P \prior{a}\\
\setleft{\mbox{So,}} \tilde{a}&=\inv{\left(\zeta 
D+P\right)}\left(\m+\zeta D\prior{a}\right)
\>
A modified expression for $\rho_{\m} (\j)$ is now available,
\<
\rho_{\m} (\j)
& \propto  \exp \left\{ 
    -\frac1{2\alpha^2} 
    \trn{\col{a-\tilde{a}}{b-\prior{b}}} \diag{P\inv{D}P+\zeta P}{\zeta Q} 
    \col{a-\tilde{a}}{b-\prior{b}}
  \right\} \\
\label{eqn:aposteriori}
& \propto  \exp \left\{ 
    -\frac1{2\alpha^2} 
    \trn{\col{a-\tilde{a}}{b-\prior{b}}} 
   \inv{\diag{\inv{\left(P\inv{D}P+\zeta P\right)}}{\inv{Q}/\zeta}}
    \col{a-\tilde{a}}{b-\prior{b}}
  \right\}
\>
This is explicitly in the form of a gaussian from which the mean current
and covariance matrix can be identified by inspection. At this stage it is
clear that the mean value of $b$ is $\prior{b}$ i.e. that there is no change
from our prior knowledge (this is because by construction all the information
that the experiment provides is orthogonal to the $\li$). 

In order to produce images of the {\em a posteriori} current density, it is 
necessary to find the distribution of a single statistic that can be
computed. The statistic is defined through a `test current' $\t = \chi_{V_k}
\ofr
\hat{e}_{\alpha}$ where $\chi_{V_k} \ofr$ is the characteristic function of a
voxel in the brain and $\hat{e}_{\alpha}$ is a unit vector. This is another 
departure from Clarke \cite{e:Clarke94} in which a delta function test current
is assumed. This choice causes problems because the inner product 
$\ip{\t}{\t}$ which is needed below
(see Equation~\ref{eqn:variance}) is undefined for a delta function.

The distribution of the statistic $\lambda=\ip{\t}{\j}$ will now be determined.
First, the coefficients of the basis elements required to 
construct \t\ are identified: 
\<
\lambda & =  \ip{\t}{\j} \\
& =  \ip{\t}{\sum_i a_i \omega \ofr \Li \ofr +\sum_i b_i \li \ofr} \\
& =  \sum_i a_i \ip{\t}{\omega \ofr \Li \ofr} +\sum_i b_i \ip{\t}{\li \ofr} \\
& =  \trn{u} a+ \trn{v}b \mbox{, say.}
\>

Equation~\ref{eqn:aposteriori} is projected onto this particular 
linear combination of co-ordinates to find the probability density of 
$\lambda$
\[
\label{eqn:rhom}
\rho_{\m} (\lambda)  \propto \exp \left\{ 
    -\frac1{2\alpha^2}
    \frac{\left(\trn{u}a+\trn{v}b-\trn{u}\tilde{a}-\trn{v}\prior{b}\right)^2}
    {\trn{u}\inv{\left(P\inv{D}P+\zeta P\right)}u +\trn{v}\inv{Q}v/\zeta}
  \right\}
\]

The mean of $\lambda$ can be identified from Equation~\ref{eqn:rhom} by 
inspection.
\[
\label{eqn:meanish}
\mbox{mean of $\lambda$} = \trn{u}\tilde{a}+\trn{v}\prior{b}
\]
The term $\trn{v}\prior{b}$ cannot be computed directly since the basis
fields $\li$ have not been defined explicitly. The problem may be overcome by
expanding $\ip{\t}{\jp}$
\<
\ip{\t}{\jp}
& =  \ip{\t}{\sum_i \prior{a}_i \omega \ofr \Li \ofr
     +\sum_i \prior{b}_i \li \ofr}\\
& =  \sum_i \prior{a}_i \ip{\t}{\omega \ofr \Li \ofr}
     +\sum_i \prior{b}_i \ip{\t}{\li \ofr}\\
& =  \trn{u}\prior{a}+\trn{v}\prior{b}
\>
Equation~\ref{eqn:meanish} may now be rewritten using only references to 
known vector fields as:
\<
\mbox{mean of $\lambda$} 
& = \trn{u}\tilde{a}+\ip{\t}{\jp}-\trn{u}\prior{a}\\
&= \trn{u} \inv{\left(\zeta D+P\right)}\left(\m+\zeta D\prior{a}\right)
+\ip{\t}{\jp}-\trn{u}\prior{a}\\
\label{eqn:mean}
&= \trn{u} \inv{\left(\zeta D+P\right)}\left(\m-P\prior{a}\right)
+\ip{\t}{\jp}
\>

This equation explicitly writes the expectation value of the statistic
$\lambda=\ip{\t}{\j}$ as a sum of two terms. The second term is the statistic
for the prior current, and so the first term can be identified as the 
correction to the prior suggested by the measurements i.e. the first term shows
the difference between the expectation of $\ip{\t}{\j}$ before and after the 
experiment was made. This is the first central result of this paper 
and it is worth stating explicitly:
\[
\label{eqn:difference}
\mbox{Change in expectation of $\ip{\t}{\phantom{\j}}$} 
= \trn{u} \inv{\left(\zeta D+P\right)}\left(\m-z(\jp)\right)
\]

Using Equation~\ref{eqn:rhom} it is possible go further than this and determine 
the statistical significance of the statistic. This is because the variance
of the 
variable $\lambda$ can also be read off from Equation~\ref{eqn:rhom} as:
\[
\label{eqn:nearlyvariance}
\mbox{variance of $\lambda$} = \alpha^2 \left[
    \trn{u}\inv{\left(P\inv{D}P+\zeta P\right)}u 
   + \frac1{\zeta} \trn{v}\inv{Q}v
  \right]
\]

In order to derive an expression in the form of computable matrices, the term 
$\trn{v}\inv{Q}v$ must be rewritten. To do so, \t\ is written as a linear 
combination of basis elements:
\[
\t \ofr = \sum_i x_i \omega \ofr \Li \ofr + \sum_i y_i \li \ofr
\]
Of course $x$ and $y$ are related to $u$ and $v$. In fact:
\<
u_i 
& = \ip{\t}{\omega \ofr \Li \ofr}\\
& = \ip{\sum_j x_j \omega \ofr \Lj \ofr + \sum_j y_j \lj \ofr}
  {\omega \ofr \Li \ofr}\\
& = \sum_j x_j \ip{\omega \ofr \Lj \ofr}  {\omega \ofr \Li \ofr}\\
& = \sum_j x_j  P_{ij}
\>
Similarly, $v=Qy$. Using these relationships, 
the inner product of \t\ with itself can be computed:
\<
\ip{\t}{\t}
& =  \ip{\sum_i x_i \omega \ofr \Li \ofr + \sum_i y_i \li \ofr}
  {\sum_j x_j \omega \ofr \Lj \ofr + \sum_j y_j \lj \ofr} \\
& =  \trn{x}Px+\trn{y}Qy \\
\>

from which, 
\<
\label{eqn:ipt}
\ip{\t}{\t}& =  \trn{u}\inv{P}u+\trn{v}\inv{Q}v
\>

Equations~\ref{eqn:nearlyvariance} and~\ref{eqn:ipt} can now be combined to 
generate the following formula for the variance:
\[
\label{eqn:variance}
\mbox{variance of $\lambda$} = \alpha^2 \left[
    \trn{u}\inv{\left(P\inv{D}P+\zeta P\right)}u 
   + \frac1{\zeta} \left( \ip{\t}{\t} - \trn{u}\inv{P}u\right)
  \right]
\]

This equation is a generalisation of the results of  Clarke \cite{e:Clarke94},
some consequences
of it were explored in \cite{p:Hasson93C} and \cite{p:Hasson95E}. It is 
interesting to note that the existence of a prior current density does not 
affect this variance.

It may be helpful to relate features of this equation to the measurement
system. 
The second term in Equation~\ref{eqn:variance} is multiplied by 
$\alpha^2/\zeta=\beta^2$ and so is independent of the assumed noise levels in
the detectors. It represents a variance derived from the finite number of 
measurements and the geometry of the experiment. Since it is
proportional to $\beta^2$, it can be interpreted in terms of the 
truncation error becoming less and less important as the certainty of the
prior distribution \jp\ increases. 

The first term in Equation~\ref{eqn:variance} is proportional to $\alpha^2$.
It shows how the noise in the data is reflected into source space. The
unregularised form of this term was derived previously by Ioannides et al in
\cite{e:Ioannides90} using an {\it ad hoc} argument. Ioannides et al obtained 
the regularised form by replacing occurrences of 
$\inv{P}$ in the unregularised form by $\inv{(P+\zeta I)}$. In the notation 
used here this gives a first term as follows
\[
\mbox{first term} = 
\trn{u}\inv{\left(P+\zeta I\right)}\inv{\left(P+\zeta I\right)}u 
\]
This does not agree with Equation~\ref{eqn:variance} which has the form (in 
the case of the measurement error being uncorrelated, i.e. $D=I$) 
\[
\mbox{first term} = 
\trn{u}\inv{P}\inv{\left(P+\zeta I\right)}u 
\]

The behaviour as $\zeta$ tends to infinity is that the variance tends to zero.
This is reasonable since, for fixed experimental noise levels (i.e. fixed
$\alpha$), $\zeta$ tending to infinity corresponds to $\beta$ tending to zero, 
which in turn corresponds to greater and greater certainty that the prior is 
correct. When $\beta$ is zero the {\em a priori} current distribution is known 
with absolute certainty. This is consistent with the above analysis which
indicates that, in this case,  the {\em a posteriori} current density is
certain to be equal
to the {\em a priori} current density. 

Note that, when computing the term $\trn{u}\inv{P}u$ in 
Equation~\ref{eqn:variance}, any reasonable algorithm (e.g. Choleski's
algorithm) for computing $\inv{P}u$ can be used even though the matrix $P$ 
is ill-conditioned. This is because the large residual vector which results in
this calculation is annihilated by the inner product with $u$. So the
computation of the whole term is well conditioned. 

\section{Applications}

The main analytical results of this paper (Equations \ref{eqn:mean},
\ref{eqn:difference} and \ref{eqn:variance}) provide
the means of solving the MEG inverse problem with specific assumptions and
of assessing the robustness of the solution.
In this section, this approach will be illustrated through three studies: a
simulation of a few-dipole source set; an analysis of the appropriate dipole
model for data from a real experiment on face processing; and a comparison of
responses to different visual stimuli from the same real experiment. 

\begin{figure}[htbp]
\begin{center}
\begin{picture}(165,50)
\put(0,10){\mbox{\epsfxsize=48mm \epsffile{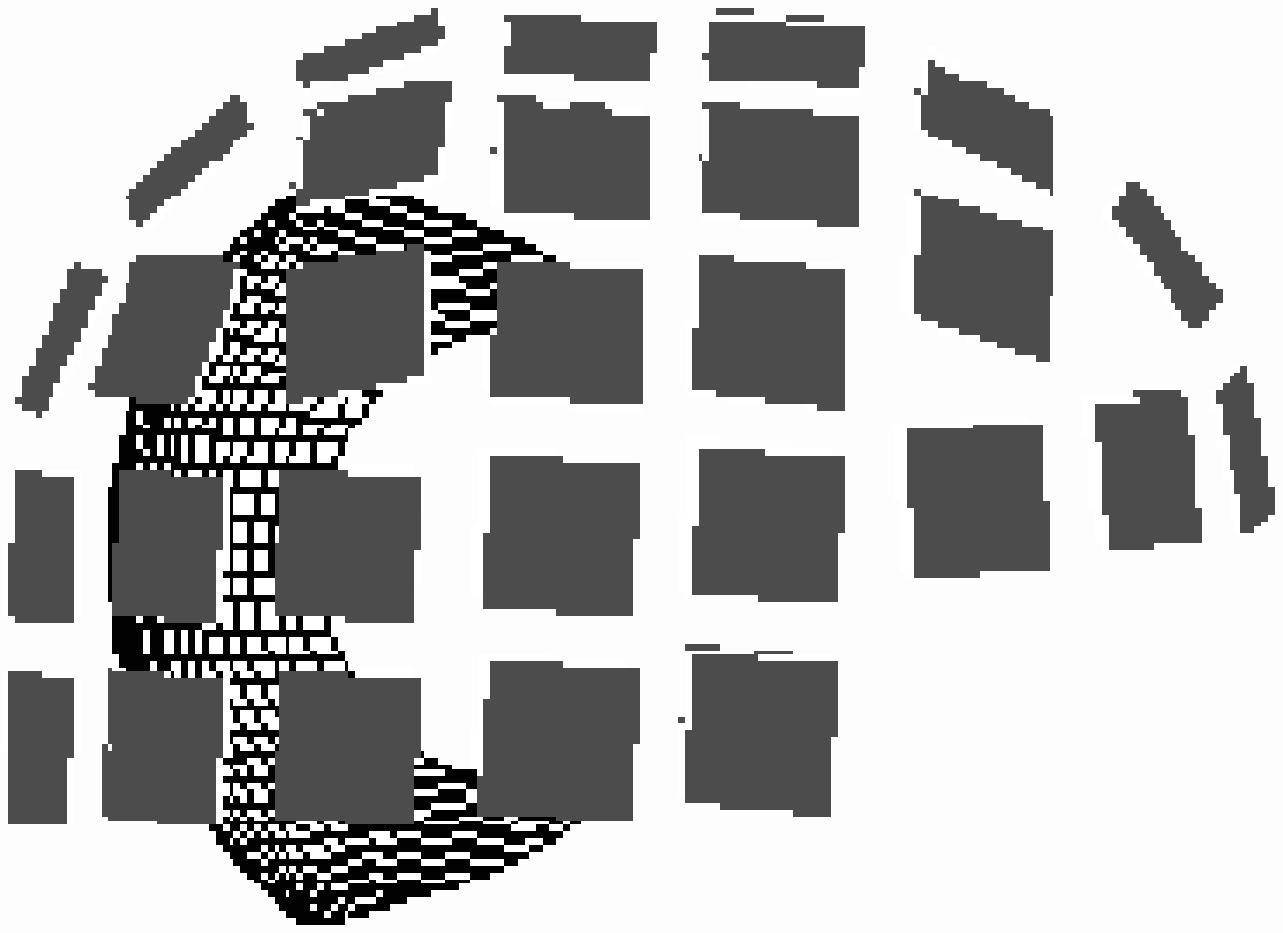}}}
\put(110,7){\mbox{\epsfxsize=48mm \epsffile{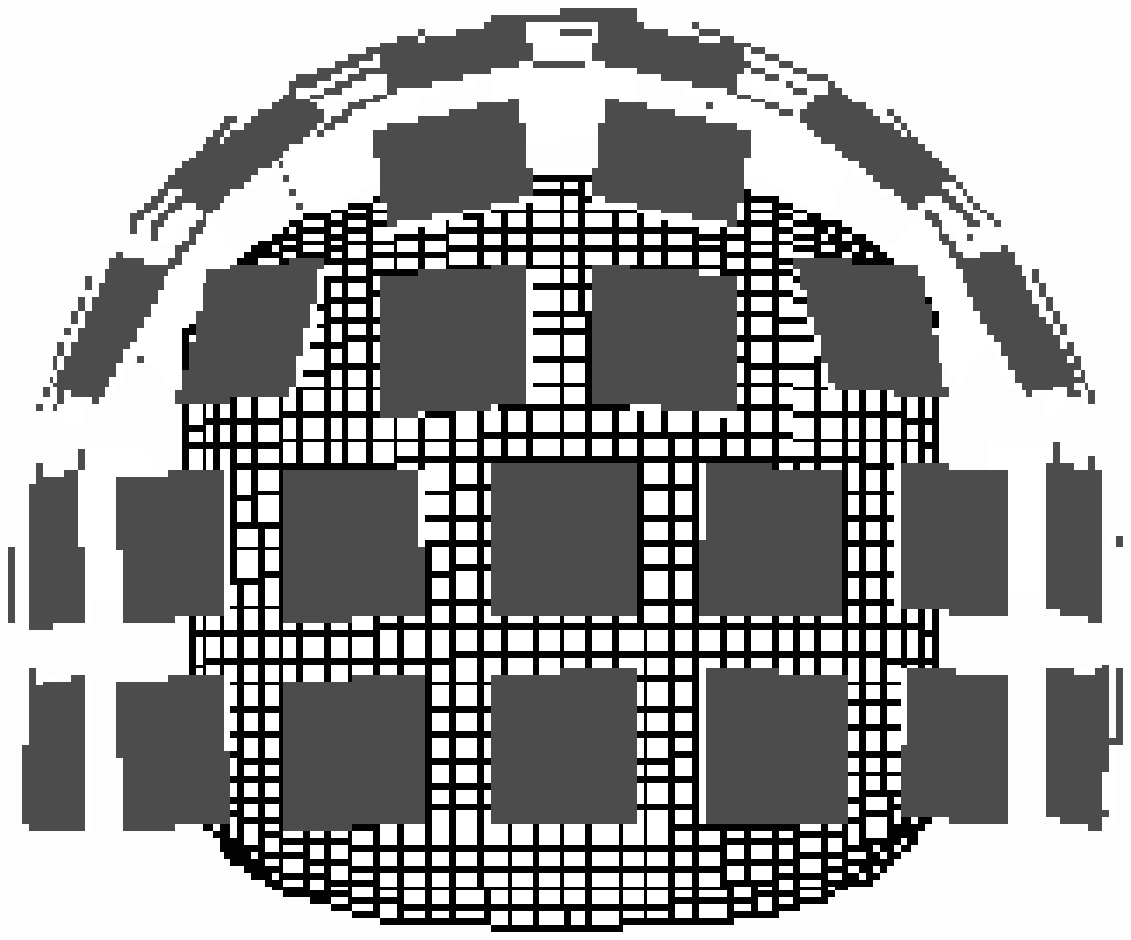}}}
\put(55,0){\mbox{\epsfysize=48mm \epsffile{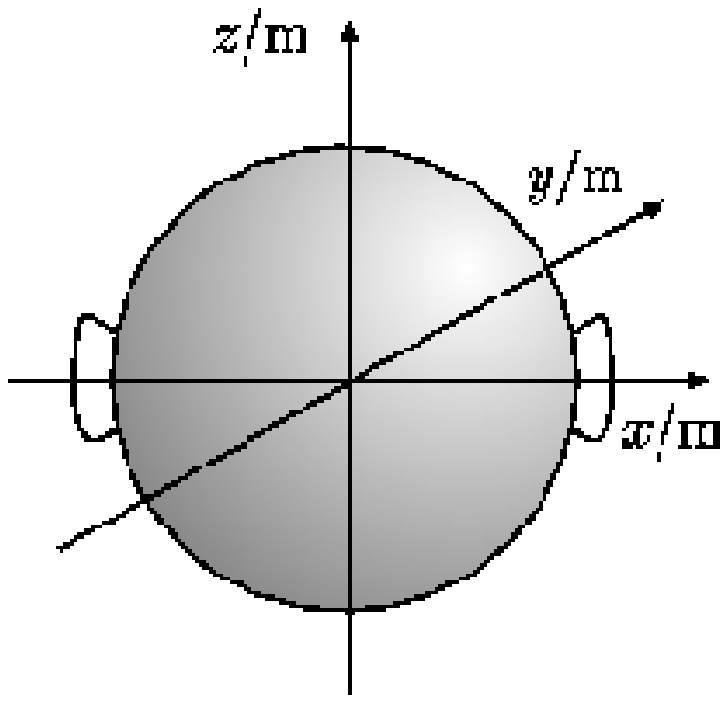}}}
\put(0,45){\mbox{(a)}}
\put(55,45){\mbox{(b)}}
\put(110,45){\mbox{(c)}}
\end{picture}
\end{center}
\caption{\label{fig:geom}
The experimental geometry. a) and c) show the 61 measurement sites in
the helmet arrangement. Each square represents a pair of orthogonal
detectors. Also shown is the source space, consisting of a discretised
mesh covering a part of a spherical shell. b) Shows the coordinate system with
the x-axis along the line joining the pre-auricular points and the
y-axis joining the inion and nasion.}
\end{figure}

All the illustrations are based on the same experimental arrangement and the same instrument, the Neuromag-122$^{TM}$ 
\cite{e:Knuutila93}. This is a helmet MEG system that contains 61 pairs of
first order gradiometers ($\partial B_r/\partial\theta$, 
$\partial B_r/\partial\phi$
in spherical polar co-ordinates), covering the head
(Figure~\ref{fig:geom}). The outputs of each pair of detectors are
closely related to the dominant tangential ionic current flow in the region
underlying the relevant sensors. Also shown in Figure~\ref{fig:geom} is the 
assumed source space, a 2-d spherical shell of radius  0.08\,m covering
a 2 radian by 2 radian solid angle over posterior regions of the cortex.

\begin{figure}[htbp]
\begin{picture}(165,105)
\put(0,65){\mbox{\epsfxsize=48mm \epsffile{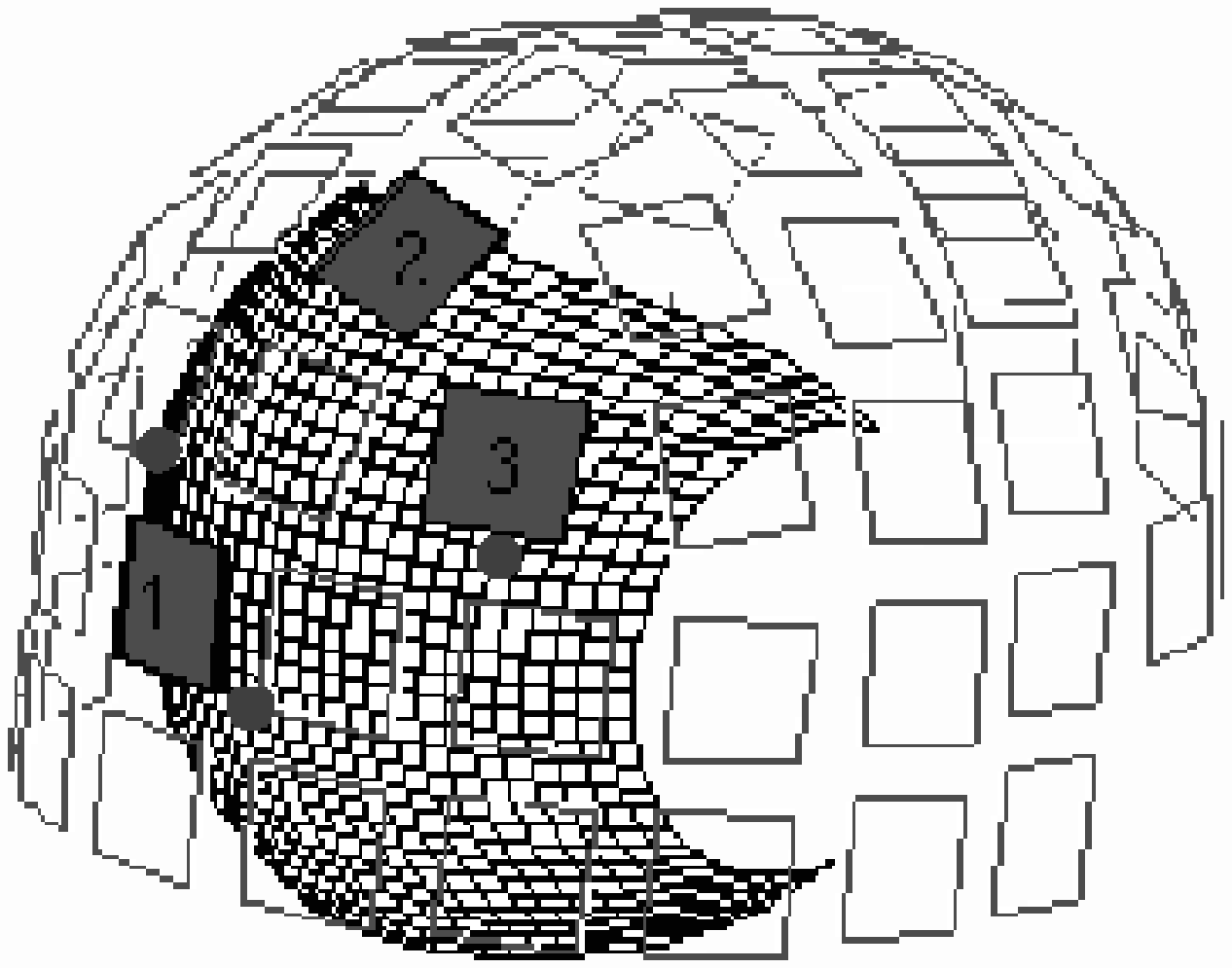}}}
\put(55,55){\mbox{\epsfysize=48mm \epsffile{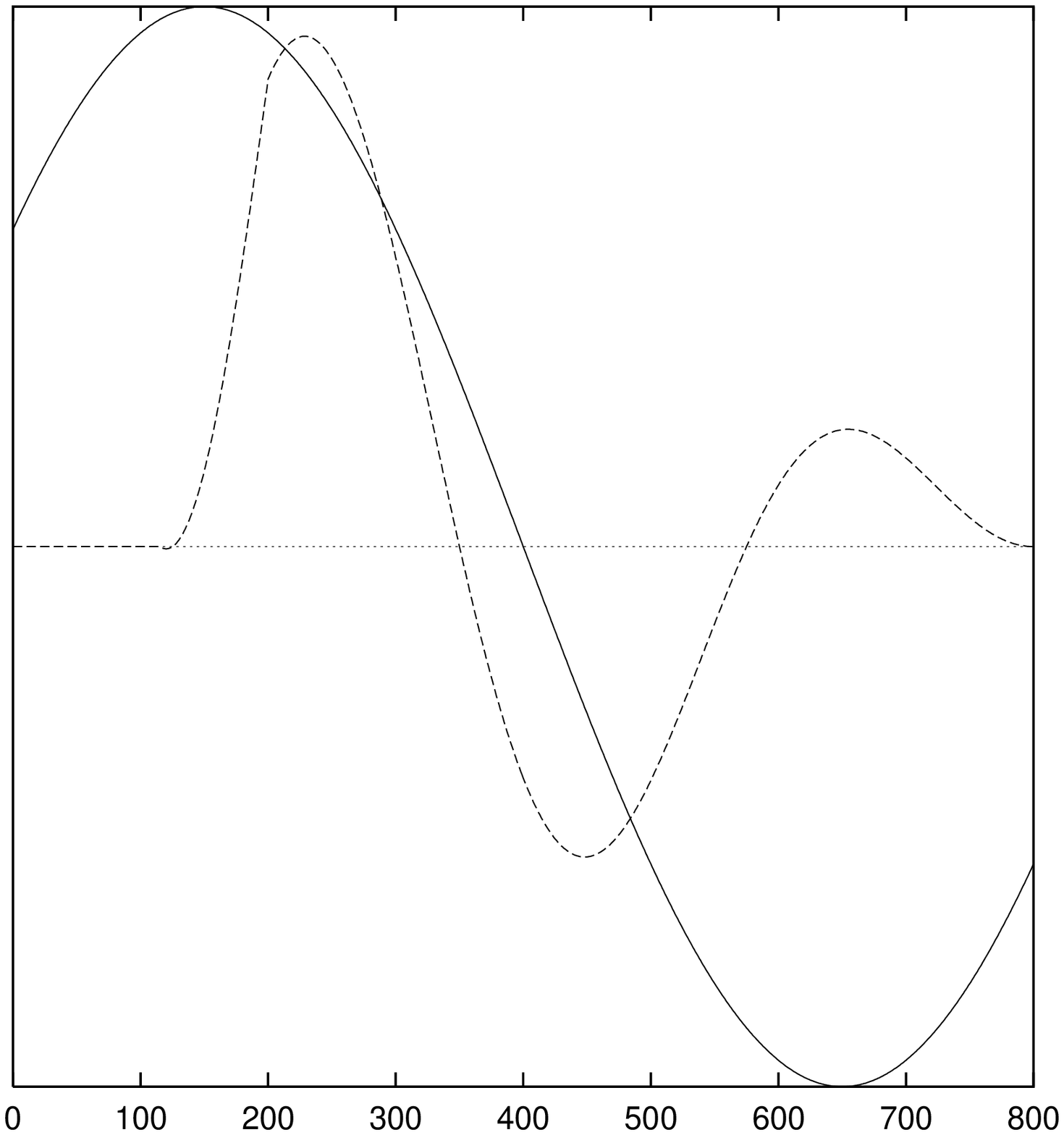}}}
\put(110,55){\mbox{\epsfxsize=48mm \epsffile{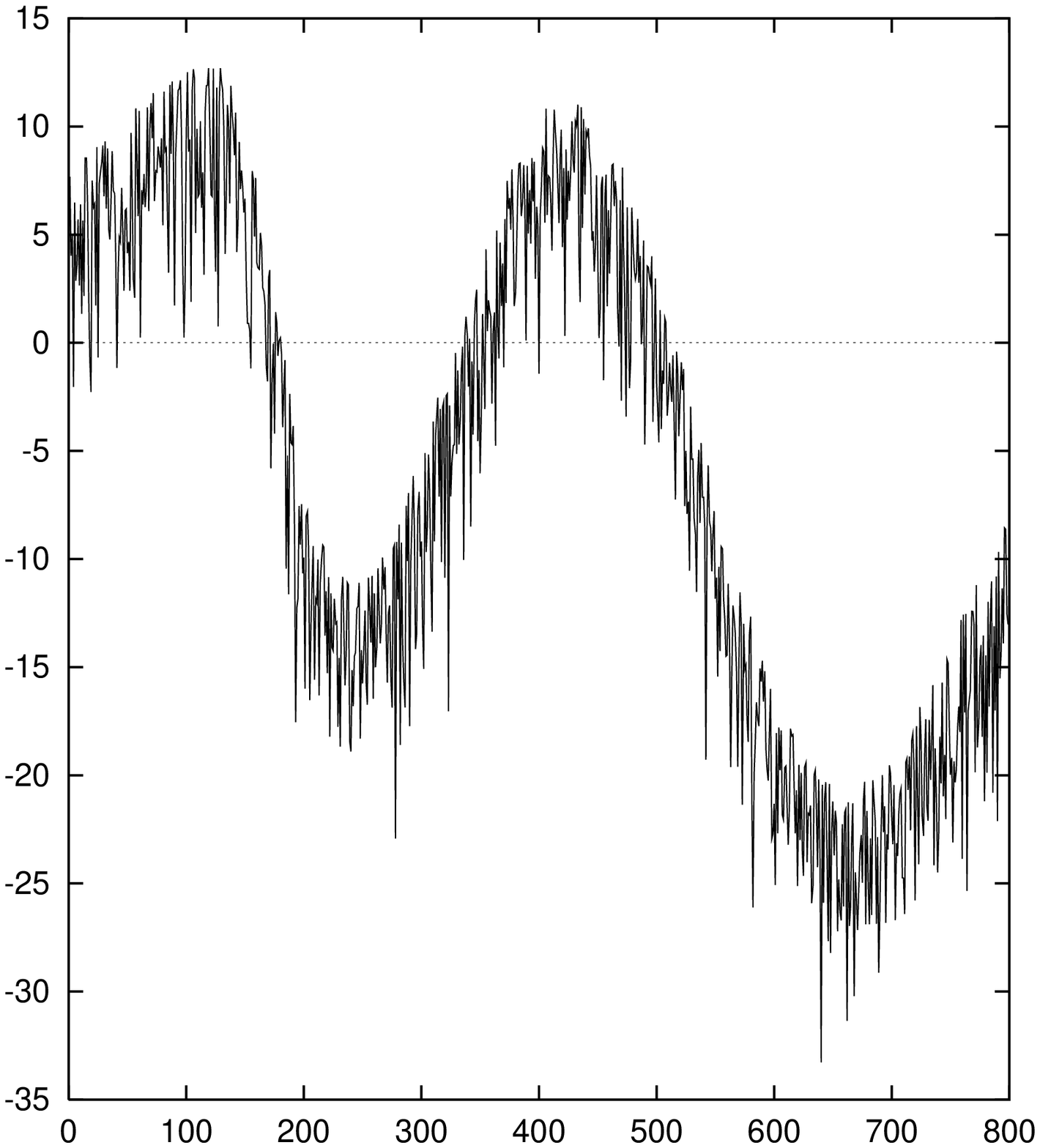}}}
\put(0,100){\mbox{(a)}}
\put(55,100){\mbox{(b)}}
\put(110,100){\mbox{(c)}}
\put(18,0){\mbox{\epsfysize=48mm \epsffile{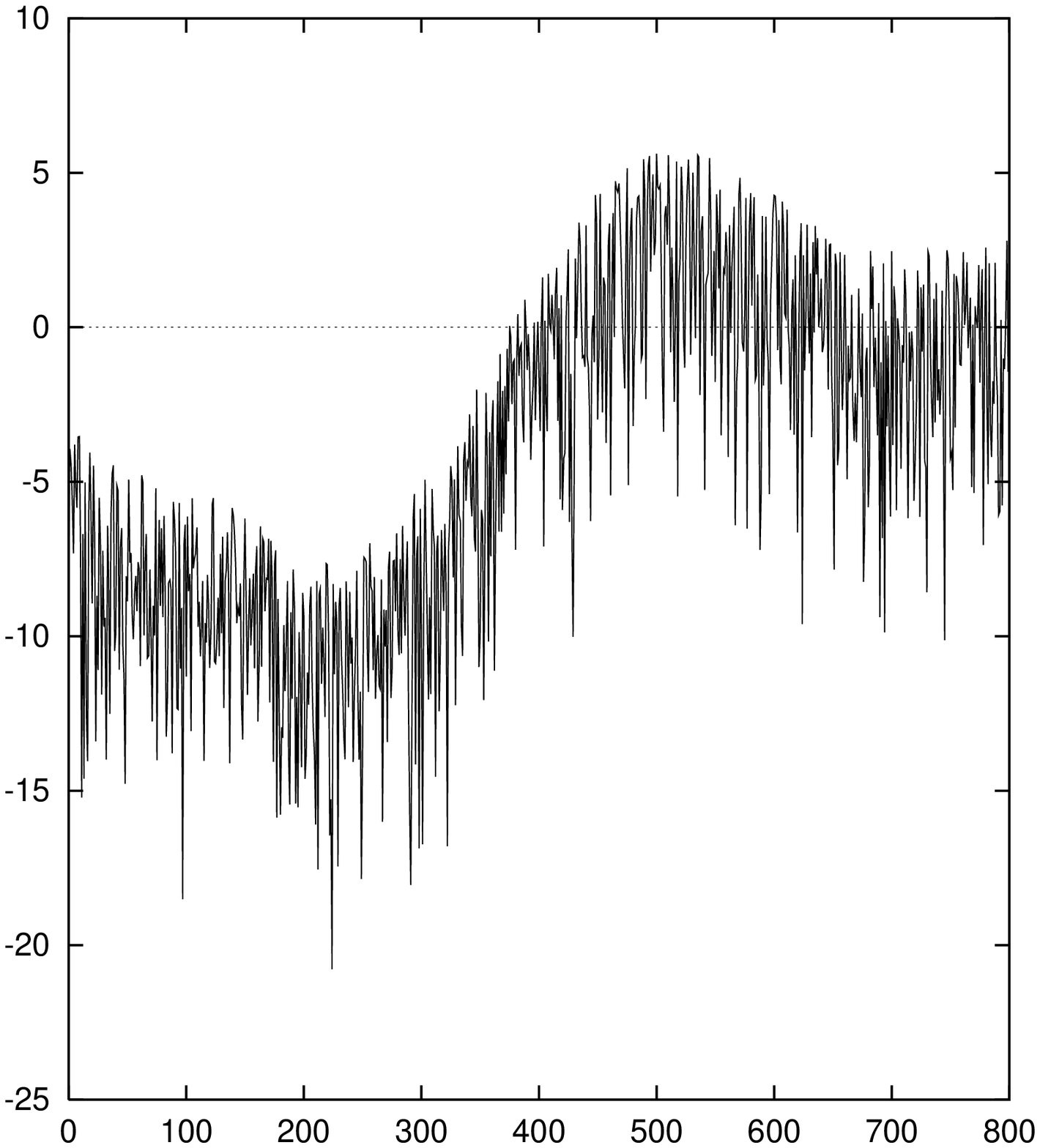}}}
\put(91,0){\mbox{\epsfxsize=48mm \epsffile{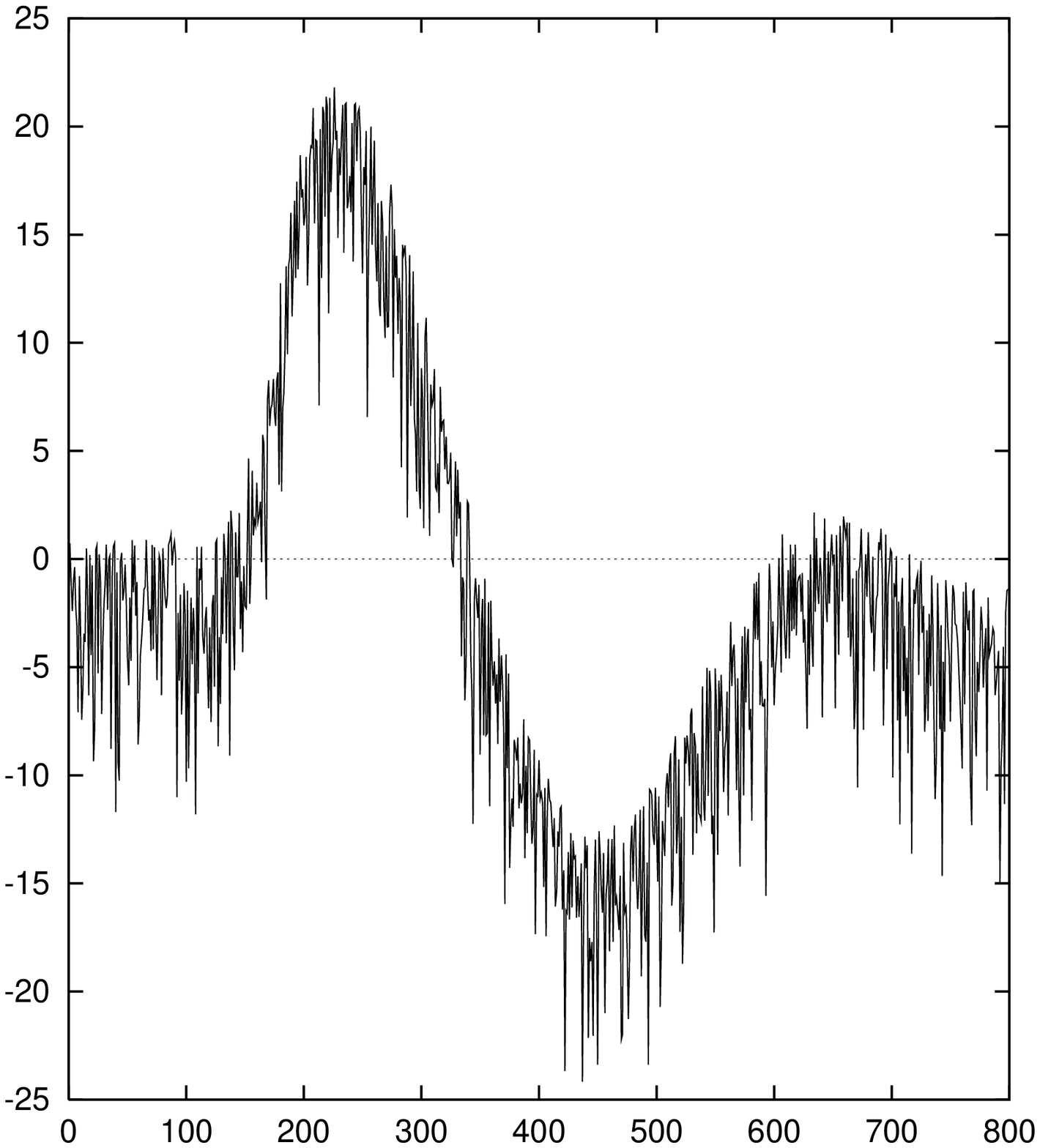}}}
\put(18,45){\mbox{(d)}}
\put(91,45){\mbox{(e)}}
\end{picture}
\vspace{1ex}
\caption{\label{fig:actcd}
a) An isometric projection of the experimental geometry. The dots on the
source space are the feet of the perpendiculars from the three dipole
sources used in this simulation. The activation curves (dipole moment vs.
time in milliseconds) for these three dipoles are shown in b). The longer
period curve corresponds to the dominant (central) source. The other curve
corresponds to the synchronous and equally activated lateral sources. The
highlighted detector sites are those for which the data is shown in c) channel
1 d) channel 2 and e) channel 3. Note that it is only in the case of Channel 3
that the signal follows clearly the activation curve for the closest dipole. }
\end{figure}

The source configuration for the first simulation study was
three dipoles distributed in an isosceles triangle configuration (co-ordinates
(0,-0.08\,m,-0.02\,m), (-0.05\,m,-0.06\,m,0.02\,m) and 
(0.05\,m,-0.06\,m,0.02\,m) with orientations
(1,0,1), (1,0,-1), (1,0,1) respectively). 
The positions of the dipoles were chosen so as
to represent approximately a central primary and two bilateral secondary
visual processing areas. The precise locations are not exactly on the 
source space shell but are displaced by 3\,mm, 1\,mm,
and 1\,mm respectively from the shell. The dipole locations and activation curves
are shown in Figure~\ref{fig:actcd}. The central source is activated first,
followed by synchronous activation of the lateral sources. The forward problem
is solved using a homogeneous sphere conductor model centred at the origin.
Gaussian noise has been
added to the computed dipole signal so that the integrated noise power is
equal to 50\% of the integrated signal power. Examples of simulated data are
shown in Figure~\ref{fig:actcd}c. 

\begin{figure}[htbp]
\begin{picture}(165,2)
\put(0,-3){\mbox{(a)}}
\put(55,-3){\mbox{(b)}}
\put(110,-3){\mbox{(c)}}
\end{picture}
\parbox{\textwidth}{
\hfill \mbox{\epsfysize=48mm \epsffile{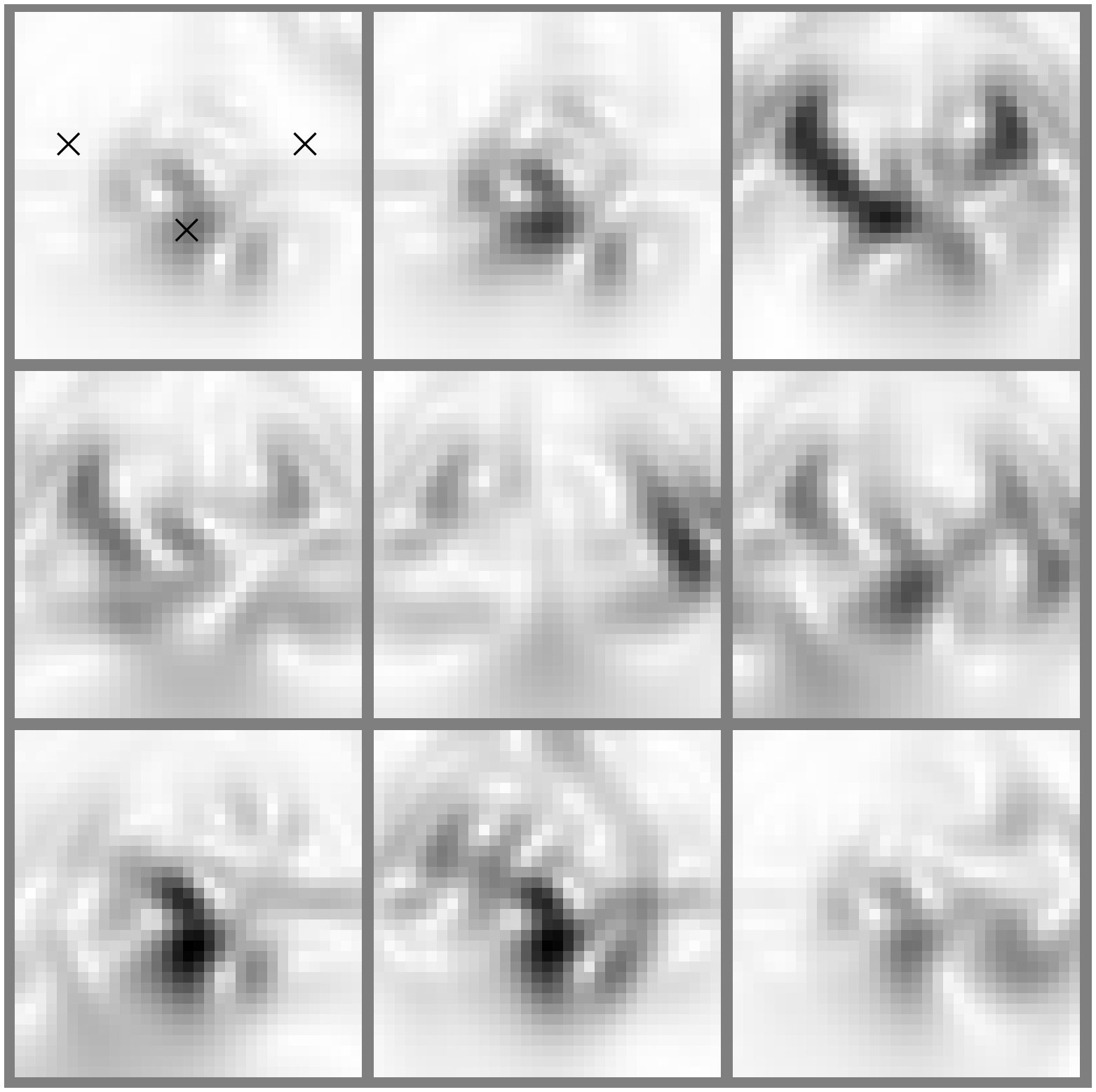}}
\hfill \mbox{\epsfysize=48mm \epsffile{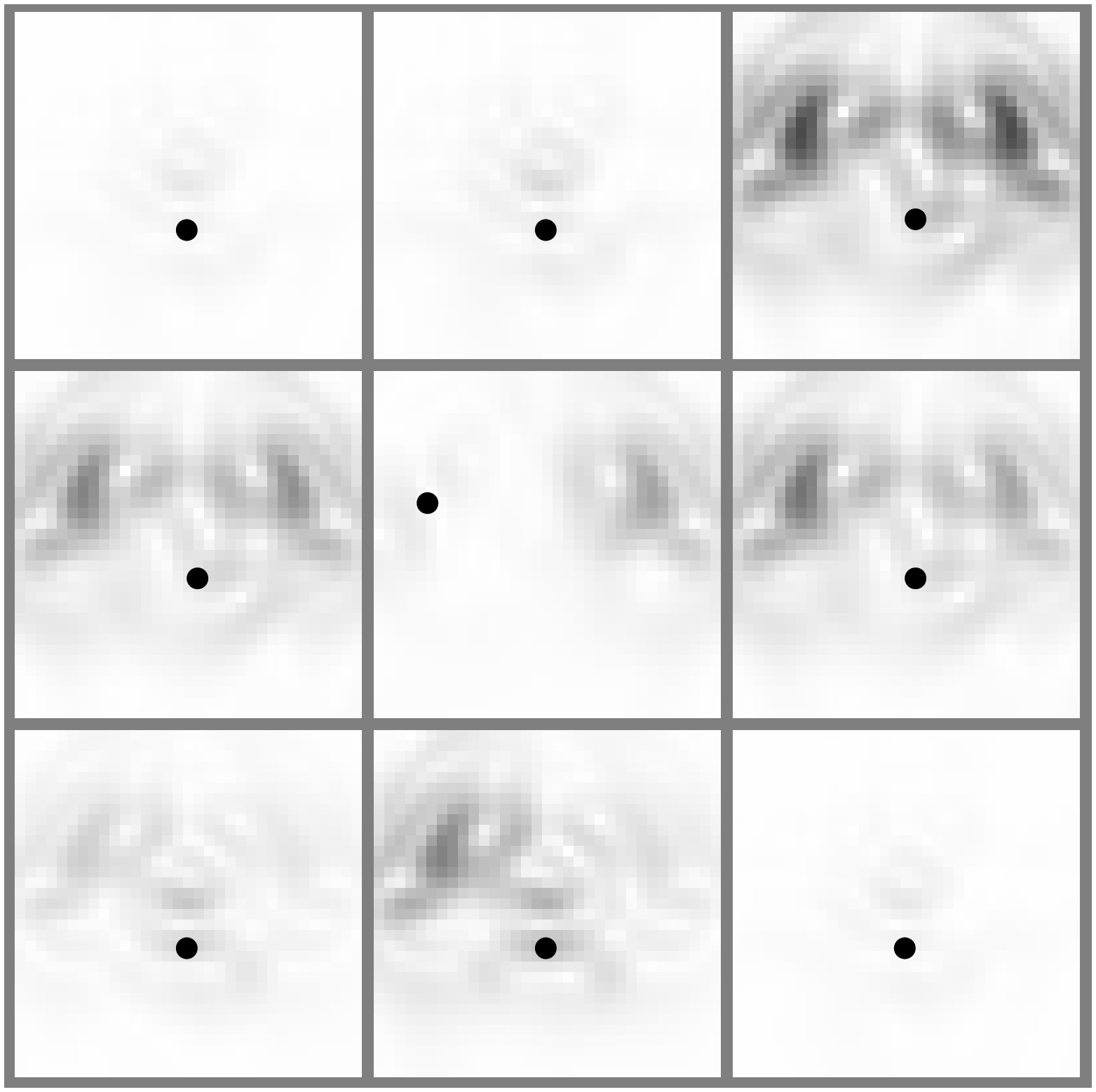}}
\hfill \mbox{\epsfysize=48mm \epsffile{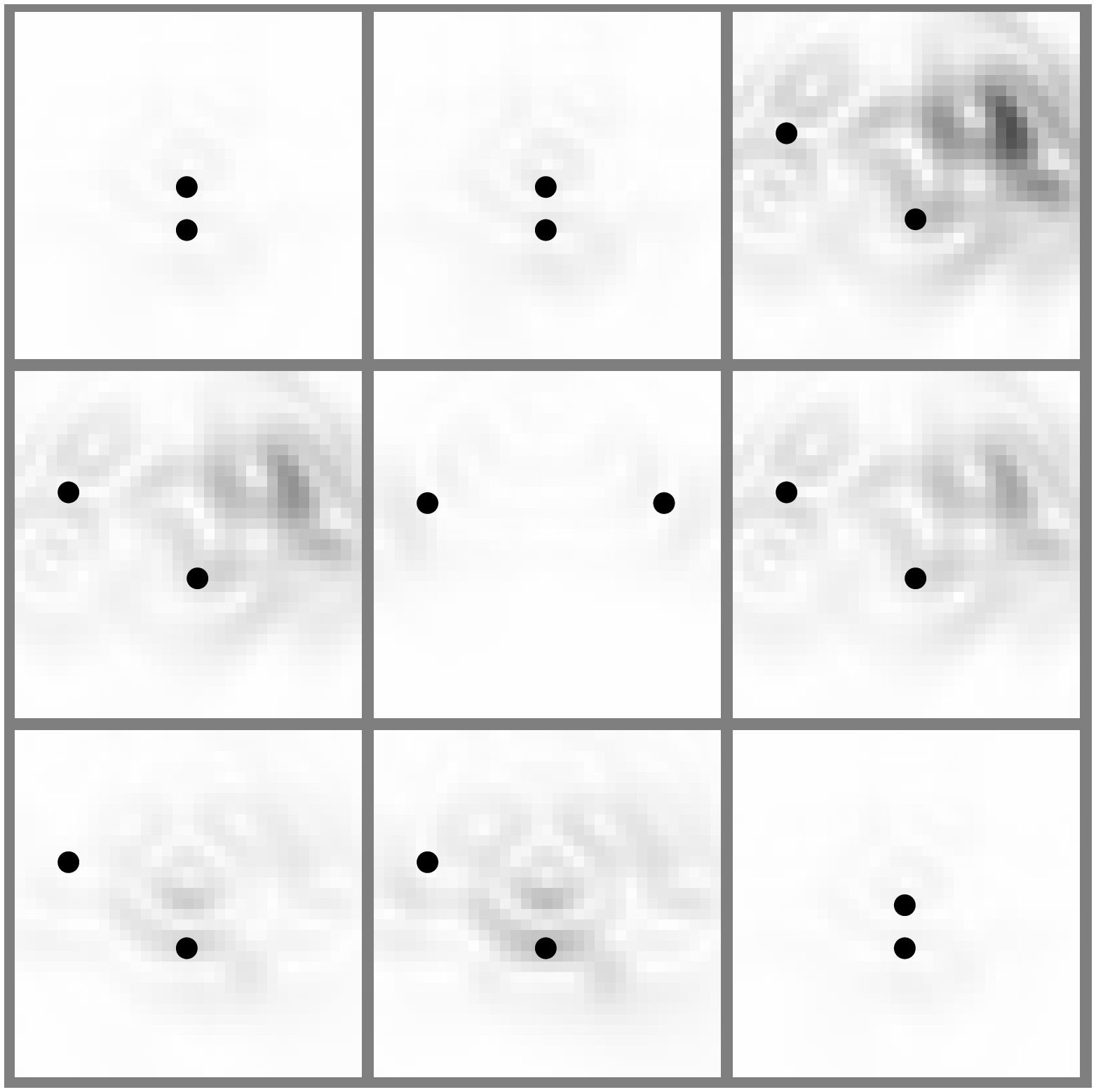}}
\hfill
}
\vspace{1ex}
\caption{\label{fig:modelsig1d} (a) The maximum likelihood current
density with a zero {\em a priori}  current distribution. 
Each frame is a two dimensional representation of the shell 
source space pictured in Figure~2. 
The 9 frames are snapshots at time 0\,ms, 100\,ms, 200\,ms, etc.
(see Figure 3). They should be read 
left to right and top to bottom. 
The optimal regularization parameter was determined by the L-curve method
\cite{e:Hansen94} to be $0.283\times trace(P)/N$. 
The crosses on the first frame of this sequence show the positions of the three
source dipoles in this representation of the source space.
(b) The significant differences between a 
single moving dipole model and the computed data. The black dots denote the 
projected position of the fitted dipole. (c) As (b) but with a two 
moving dipole model.}
\end{figure}

The simplest approach to the inverse problem is to use Equation~\ref{eqn:mean}
with a zero prior current distribution. This simplification results in the
same formulae as the probabilistic algorithm that has been used for several
years \cite{e:Clarke91,e:Clarke89}. The resulting expectation of the 
{\em a-posteriori} current distribution is shown 
in Figure~\ref{fig:modelsig1d}a. 

However, using our analysis, it is possible to employ an approach which goes
further in comparing different source descriptions. Single or few dipole
solutions can be generated from the magnetic field data and used as prior
estimates of the current distribution. It is then possible to identify the
appropriateness of this particular dipole-model prior estimate by computing
from Equation~\ref{eqn:difference} the change in the expectation associated
with including the measurement information without constraining the final
solution to a dipolar form. Because the statistic provides spatial information
it indicates directly those areas where the dipole model solution has been
modified. Using the variance associated with the {\em a-posteriori}
current (computed from Equation \ref{eqn:variance}) allows us to plot the 
number of standard deviations of the change in expectation at each point in 
source space. 

To illustrate the usefulness of this technique, two prior descriptions of the
source current for this data are postulated - a single moving dipole model and
a two moving dipole model. The optimal solutions have been found through
exhaustive search of the discretised source space by least squares
minimisation of the fit to the data. The positions of the fitted dipoles for
nine time slices are shown in Figure~\ref{fig:modelsig1d}b. They may be compared
with the nearest points on the source space to the actual dipole positions.
Using each model in turn as a prior current and Equations \ref{eqn:difference}
and \ref{eqn:variance}, the statistical significance of the differences
between the {\em a-posteriori} and the {\em a-priori} current distributions
are calculated (Figure~\ref{fig:modelsig1d}b). 

It can be seen that a single dipole is a good model for the first two time
instants and also for the ninth. This is not surprising as only one dipole is
active at these times. The significance plots for the other times suggest
systematic discrepancies between the model and the data. The restricted
localisation of these significant differences points to additional localised
sources that have been omitted from the model. The next step is the two dipole
model (Figure~\ref{fig:modelsig1d}c). The two-dipole significant difference diagram
suggests that this model is adequate for all but four time instants.
Comparison with the activation curves identifies these as being the times when
all three dipoles are active. 

The second illustration uses data from an evoked response study of
face-processing using the same experimental system 
\cite{e:Swithenby97}. Human subjects were presented briefly with
photographs of human faces and control objects 
(e.g. animals) and their neural responses were recorded as a function of time after
the stimulus. It is known that the early response to face images involves
widespread activity in the posterior brain but there is limited evidence for the
precise distribution and timecourse of the neuronal sources. One
suggestion is that there are three major areas of activity; in occipital
cortex and both right and left ventral occipito-temporal cortex 
\cite{e:Lu91,e:Halgren95}. Strong occipital activity 
(starting about 100\,ms after the
stimulus) is expected to lead to concurrent activity in the two other regions
with a stronger response in the right hemisphere \cite{e:Swithenby97}. The
hypothesised arrangement is therefore similar in geometry to the
simulated measurement already discussed. 

\begin{figure}[htbp]
\begin{picture}(165,2)
\put(0,-3){\mbox{(a)}}
\put(55,-3){\mbox{(b)}}
\put(110,-3){\mbox{(c)}}
\end{picture}
\parbox{\textwidth}{
\hfill
\mbox{\epsfysize=4.8cm \epsffile{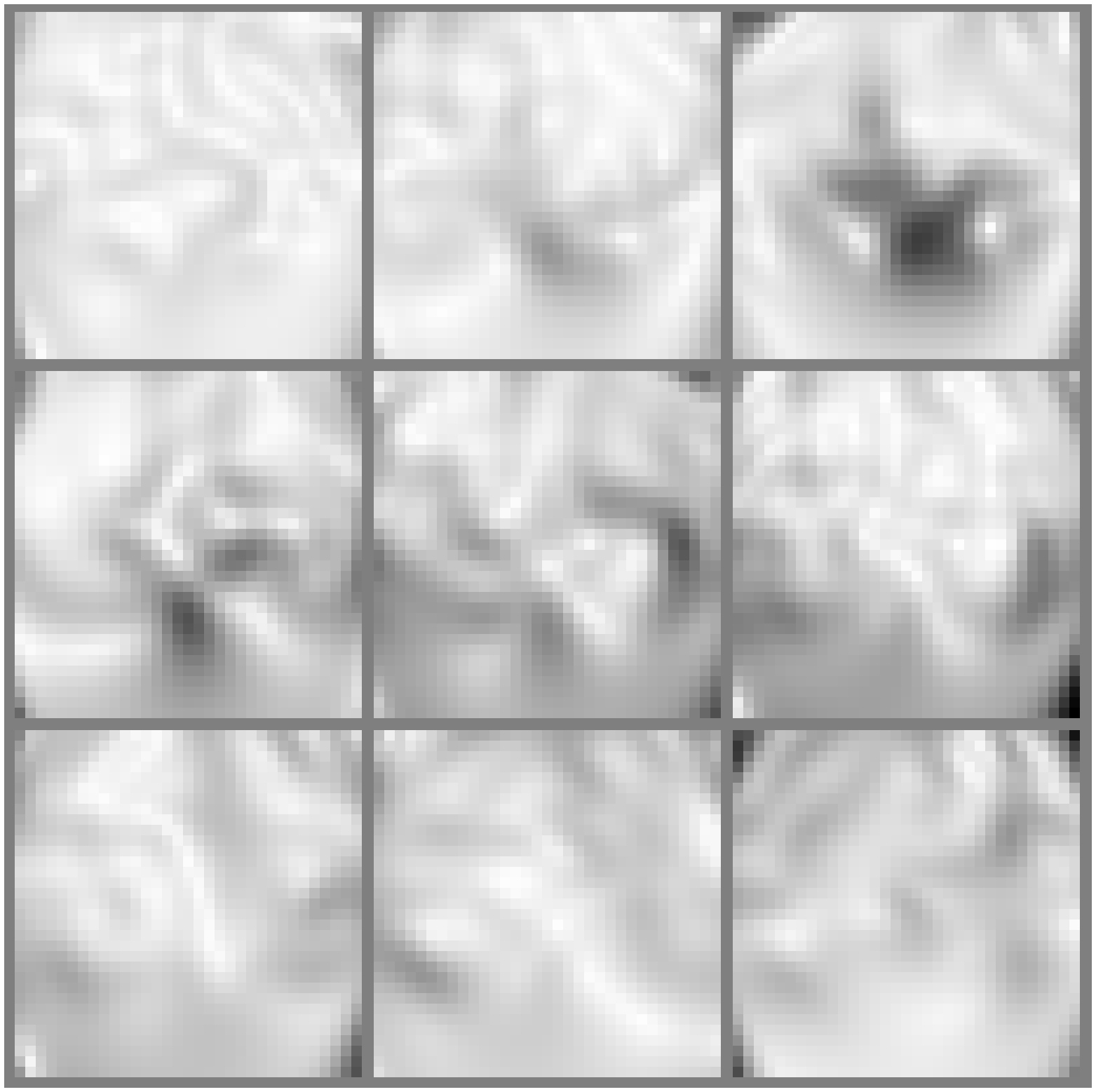}}
\hfill
\mbox{\epsfysize=4.8cm \epsffile{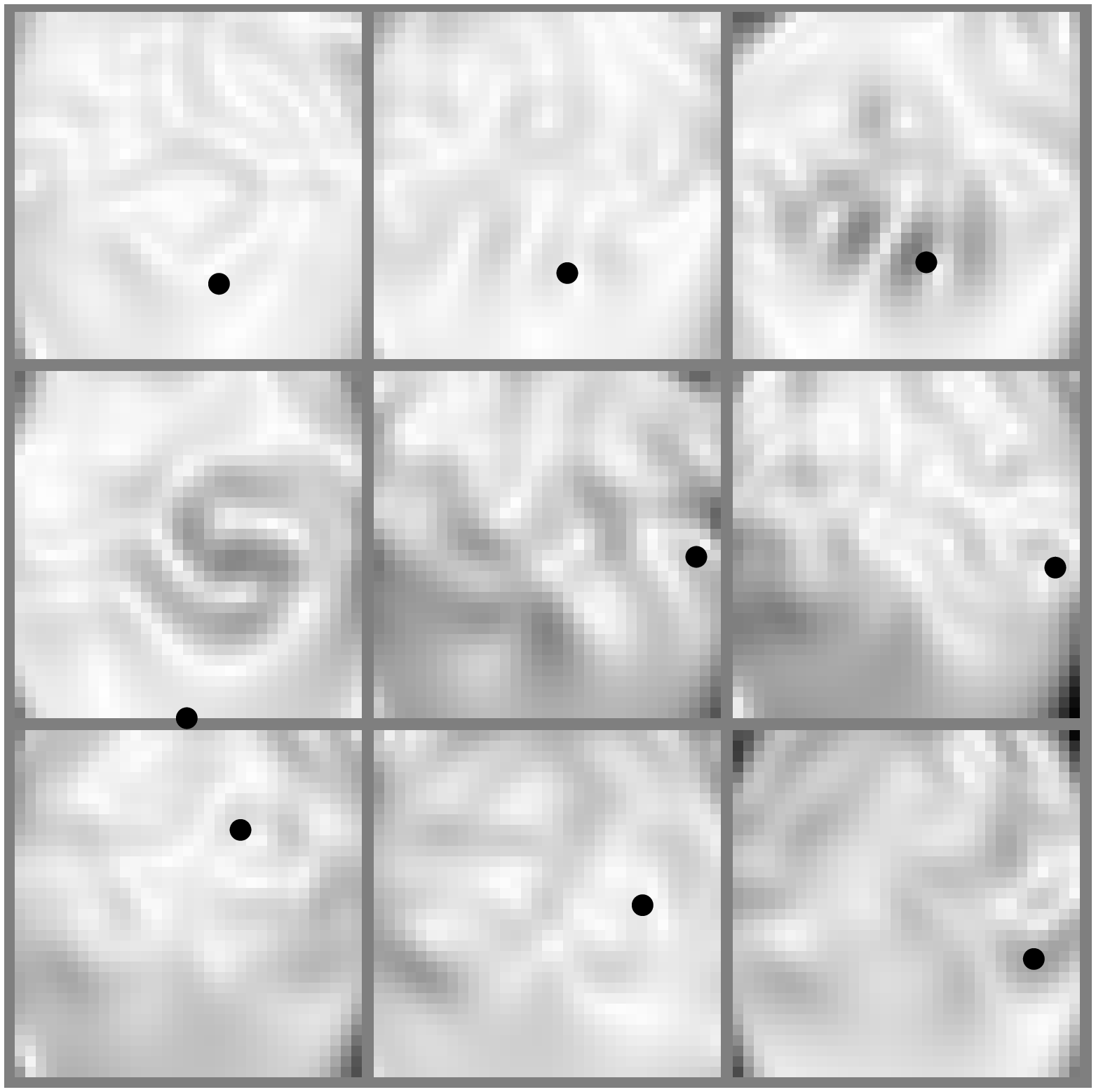}}
\hfill
\mbox{\epsfysize=4.8cm \epsffile{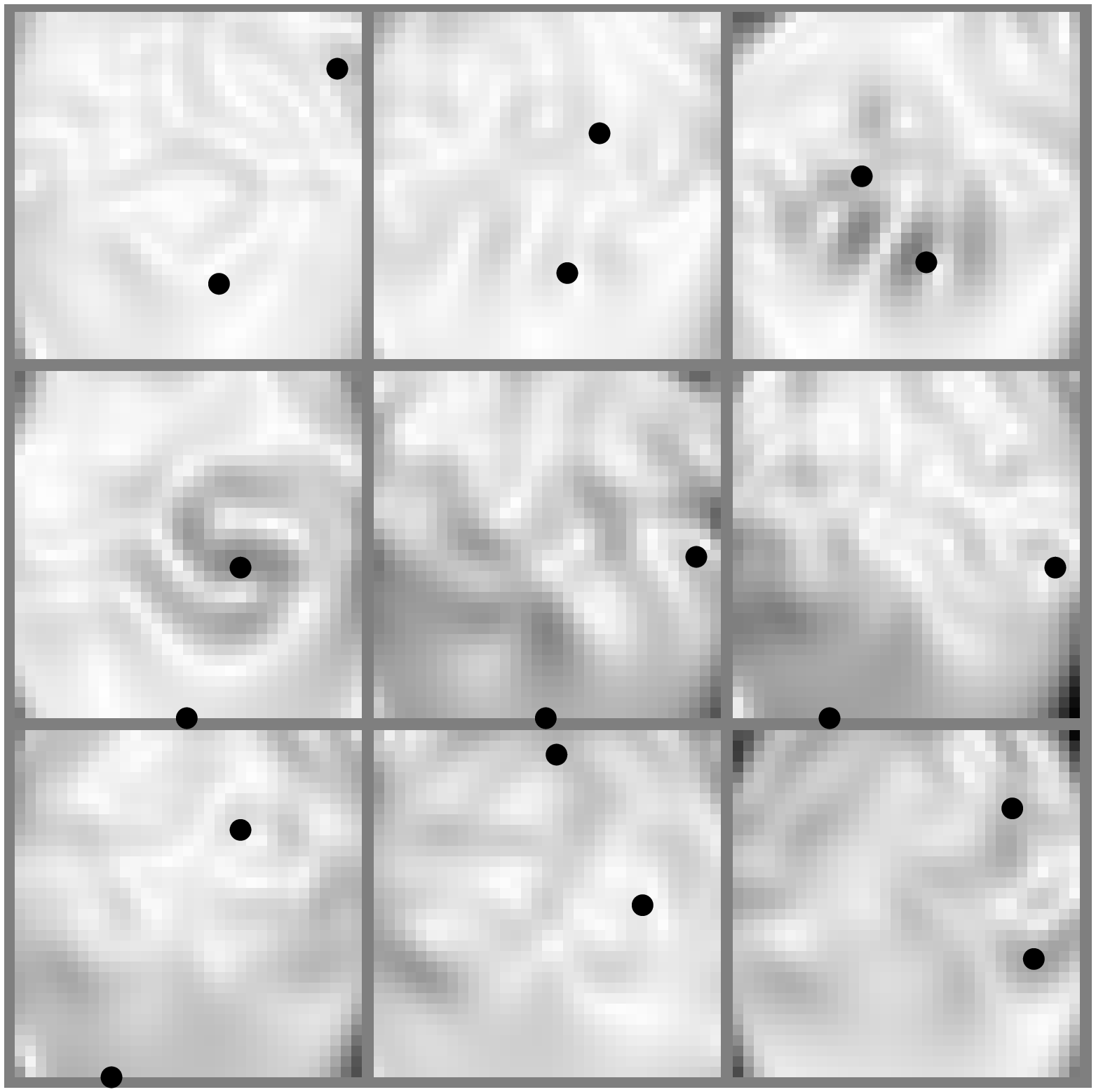}}
\hfill
}
\vspace{1ex}
\caption{\label{fig:face} (a) The maximum likelihood current density for
the face response with a zero {\em a priori}  current distribution. The 9
frames represent equal steps in time from 70\,ms to 241\,ms after the stimulus.
(b) The significant differences between a single moving dipole model
and the computed data. The black dots denote the projected positions of
a single dipole. (c) As (b) but with a two moving dipole model.}
\end{figure}

Figure~\ref{fig:face} shows the same set of outputs that were presented for
the simulated system. In this case, the fitted dipoles may be thought of as
representing a discrete limited region of source current. 
Figure~\ref{fig:face}a suggests early central activity (frame 3) followed by
less prominent localised activity on the right (frame 5). These source regions
are reflected in the 1-dipole solutions (Figure~\ref{fig:face}b). However, comparison with
Figure~\ref{fig:actcd}b shows that the accuracy of the single dipole fit is less than for
the simulated data even though the noise levels are comparable. This may
suggest that there are other active sources present. There is no evidence that
these are recovered by the 2-dipole model (Figure~\ref{fig:face}c) as there is
little improvement in the significant difference maps generated using a two 
dipole model as
the prior (see for example the strong similarity between frame 4 in
Figures~\ref{fig:face}b and~\ref{fig:face}c). It would be reasonable to infer that the additional sources are
diffuse. 

The third illustration relates to the main thrust of the face processing study
by Swithenby et al \cite{e:Swithenby97}, which was to identify statistically
significant evidence for differences between the responses to faces and the
other complex visual stimuli. In the initial analysis the strength of evoked
activity within a certain region and latency span was parameterised in terms
of the signal power integrated over a group of channels and a specified
latency span. These calculations revealed that the brain activity in the right
occipito-temporal region following face presentation is significantly
different (p=0.05) from activity following non face images during the latency
span 110 to 170 ms. No other consistent and statistically significant
differences were found. This data-space analysis, though useful, was
complicated by the need to survey the large number of possible choices of
channel group and latency range. 

\begin{figure}[htbp]
\begin{center}
\mbox{\epsfysize=4.8cm \epsffile{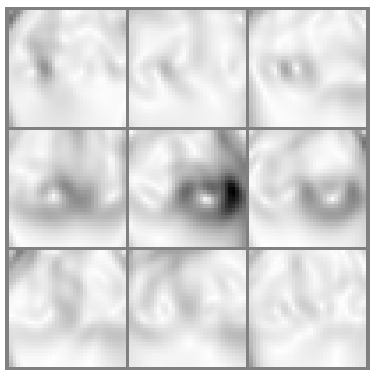}}
\end{center}
\vspace{1ex}
\caption{\label{fig:faceCompare}  The maximum likelihood current density for
the face response with a {\em a priori}  current distribution derived from the 
control experiment with pictures of animals. The 9
frames represent equal steps in time from 70\,ms to 241\,ms after the stimulus.
The regularisation parameter $\zeta$
was chosen by using the L-curve method \cite{e:Hansen94}. The optimal value
was $1.4\times trace(P)/N$. }
\end{figure}

The Bayesian framework developed here provides an alternative direct means of
directly comparing responses to two stimuli. By using one data set as the
prior and comparing it with the other data set it is possible to identify
those regions in source space where there is a statistically significant
difference between the two source structures within the same source model.
We have carried out this calculation for the face and control stimuli
as a function of position and latency for a simple 
two-dimensional source space consisting of a part spherical shell whose radius
is similar to that of the cortical surface (Figure~\ref{fig:faceCompare}). 

The evidence for statistically significant
differences in the right occipito-temporal region at about 155~ms is clear.
However there is no evidence for differences in earlier latencies, in 
particular with respect to the early source shown in Figure~\ref{fig:face}a.

\section{Discussion}

The illustrations provided above offer ways of exploiting the new
Bayesian results that have been derived. Dipole predictions have
been systematically examined using a measure that goes beyond a simple
scalar goodness of fit measure (i.e. percentage of variance accounted for)
to a statistically valid map of fitting significance. This addresses a long
standing issue, the unreliability of the goodness of fit measure as a
reliable test of the appropriateness of a given model in explaining a given
set of data \cite{e:Janday89}. The spatially discriminated
Bayesian approach gives a test which is more reliable when there is
fundamental concern about the appropriateness of a model. The second example
shows how these ideas may be applied to real data to assess the complexity of
the dipole model that a given data set can sustain. In similar fashion, the
third example illustrates how the Bayesian framework allows the direct
comparison of two data sets in order to identify quantitatively the regions of
statistically significant source activities. 

These ideas may be extended further within MEG (and EEG) data analysis. An
obvious extension is to perform dipole analysis as a precursor to a
distributed Bayesian analysis. This may provide a way of not only refining
information about the depth of source activity but also of assessing the
reliability of depth estimates. Other possibilities include exploration of the
dynamics. For example, times at which there are statistically significant
change in the data could be identified by using a source distribution
estimated from each time as the prior for an analysis of the data from the next
sampling instant.

In summary, the analysis presented here comprises a Bayesian estimator of a
source current distribution in the biomagnetic inverse problem. This is
generalisable to other systems and may be used as the engine for tests of
significant difference in source models and data.

\bibliographystyle{unsrt}
\bibliography{biomag74,biomag79,biomag82,biomag84,biomag86,biomag88,biomag90,biomag91,biomag92,personal,extra}

\begin{thebibliography}{10}

\bibitem{e:Clarke94}
C.J.S. Clarke.
\newblock Error estimates in the biomagnetic inverse problem.
\newblock {\em Inverse Problems}, 10:77--86, 1994.

\bibitem{e:Vanni96}
S.~Vanni, B.~Rockstroh, and R.~Hari.
\newblock Cortical sources of human short-latency somatosensory evoked fields
  to median and ulnar nerve stimuli.
\newblock {\em Brain Research}, 737:25--33, 1996.

\bibitem{e:Buchner95}
H.~Buchner, L.~Adams, A.~Muller, I.~Ludwig, A.~Knepper, A.~Thron, K.~Niemann,
  and M.~Scherg.
\newblock Somatotopy of human hand somatosensory-evoked potentials and {3D-NMR}
  tomography.
\newblock {\em Electroenceph. clin. Neurophys.}, 96(2):121--134, 1995.

\bibitem{e:Mauguiere97}
F.~Mauguiere, I.~Merlet, N.~Forss, S.~Vanni, V.~Jousmaki, P.~Adeleine, and
  R.~Hari.
\newblock Activation of a distributed somatosensory cortical network in the
  human brain. a dipole modelling study of magnetic fields evoked by median
  nerve stimulation. 1. location and activation timing of sef sources.
\newblock {\em Electroenceph. clin. Neurophys.}, 104(4):281--289, 1997.

\bibitem{e:Hoshiyama97}
M.~Hoshiyama, R.~Kakigi, S.~Koyama, S.~Watanabe, and M.~Shimojo.
\newblock Activity in posterior parietal cortex following somatosensory
  stimulation in man: Magnetoencephalographic study using spatio-temporal
  source analysis.
\newblock {\em Brain Topography}, 10(1):23--30, 1997.

\bibitem{Hamalainen84}
M.S. H\"am\"al\"ainen and R.J. Ilmoniemi.
\newblock Interpreting measured magnetic fields of the brain: {E}stimates of
  current distributions.
\newblock Technical Report TKK-F-A559, Helsinki University of Technology, 1984.

\bibitem{e:Hamalainen94}
M.S. H\"am\"al\"ainen and R.J. Ilmoniemi.
\newblock Interpreting magnetic fields of the brain: minimum norm estimates.
\newblock {\em Med. \& Biol. Eng. \& Comput.}, 32:35--42, 1994.

\bibitem{p:Ioannides89}
A.A. Ioannides, J.P.R. Bolton, R.~Hasson, and C.J.S. Clarke.
\newblock Localised and distributed source solutions for the biomagnetic
  inverse problem {II}.
\newblock In S.J.Williamson, M.Hoke, G.Stroink, and M.Kotani, editors, {\em
  Advances in Biomagnetism}, pages 591--595, New York, August 1989. Plenum
  Press.

\bibitem{e:Wang92}
J.Z. Wang, S.J. Williamson, and L.~Kaufman.
\newblock Magnetic source images determined by a lead-field analysis - the
  unique minimum-norm least-squares estimation.
\newblock {\em IEEE Transactions on Biomedical Engineering}, 39:665--675, 1992.

\bibitem{e:Marqui94}
R.D. Pascual-Marqui.
\newblock Low resolution brain electromagnetic tomography.
\newblock {\em Brain Topography}, 7:180, 1994.

\bibitem{e:Gorodnitsky95}
I.F. Gorodnitsky, J.S. George, and B.D. Rao.
\newblock Neuromagnetic source imaging with focuss: a recursive weighted
  minimum norm algorithm.
\newblock {\em Electroenceph. clin. Neurophys.}, 95:231--251, 1995.

\bibitem{e:ISBET-loreta}
W.~Skrandies (Ed.).
\newblock Extended discussion of {LORETA}.
\newblock {\em International Society for Brain Electromagnetic Topography
  Newsletter}, 6, ISSN 0947-5133, December 1995.

\bibitem{e:Wood98}
C.~Wood (ed.).
\newblock Workshop on the meg inverse problem.
\newblock In Advances in Biomagnetism Research: Biomag96, (Eds: C. Aine et al)
  Springer-Verlag, New York, In press, 1998.

\bibitem{e:Baillet97}
S.~Baillet and L.~Garnero.
\newblock A {B}ayesian approach to introducing anatomo-functional priors in the
  {EEG}/{MEG} inverse problem.
\newblock {\em IEEE Trans. Biomed. Eng.}, 44(5):374--385, 1997.

\bibitem{e:Schmidt97}
D.~M. Schmidt, J.S. George, and C.C. Wood.
\newblock Bayesian inference applied to the electromagnetic inverse problem.
\newblock Preprint la-ur-97-4813, Los Alamos National Laboratory, 1997.

\bibitem{p:Hasson93C}
R.~Hasson and S.J. Swithenby.
\newblock A quantitative analysis of the {EEG} and {MEG} inverse problems.
\newblock In C.~Baumgartner, L.~Deecke, G.~Stroink, and S.J. Williamson,
  editors, {\em {BIOMAGNETISM}: Fundamental research and clinical
  applications}, pages 455--457, Amsterdam, August 1993. Elsevier Science.

\bibitem{p:Hasson95E}
R.~Hasson and S.~J. Swithenby.
\newblock Aspects of non-uniqueness: how one source affects the reconstruction
  of another.
\newblock In Advances in Biomagnetism Research: Biomag96, (Eds: C. Aine et al)
  Springer-Verlag, New York, In press, 1998.

\bibitem{e:Ioannides90}
A.A. Ioannides, J.P.R. Bolton, and C.J.S. Clarke.
\newblock Continuous probabilistic solutions to the biomagnetic inverse
  problem.
\newblock {\em Inverse Problems}, 6:523--542, 1990.

\bibitem{e:Knuutila93}
J.E.T. Knuutila, A.I. Ahonen, M.S. H\"am\"al\"ainen, M.J. Kajola, P.P. Laine,
  O.V. Lounasmaa, L.T. Parkkonene, J.T.A Simola, and C.D. Tesche.
\newblock A 122-channel whole cortex {SQUID} system for measuring the brain's
  magnetic field.
\newblock {\em IEEE Trans. Mag.}, 29 (6):3315--3320, 1993.

\bibitem{e:Clarke91}
C.J.S. Clarke.
\newblock Probabilistic modelling of continuous current sources.
\newblock In J.Nenonen, H.-M.Rajala, and T.Katila, editors, {\em Biomagnetic
  Localisation and 3D modelling}, pages 117--125, Helsinki, May 1991. Helsinki
  University of Technology, Dept. of Technical Physics.

\bibitem{e:Clarke89}
C.J.S. Clarke and B.S. Janday.
\newblock The solution of the biomagnetic inverse problem by maximum
  statistical entropy.
\newblock {\em Inverse Problems}, 5:483--500, 1989.

\bibitem{e:Hansen94}
P.C. Hansen.
\newblock Regularization tools.
\newblock {\em Numerical Algorithms}, 6:1--35, 1994.

\bibitem{e:Swithenby97}
S.J. Swithenby, A.J. Bailey, S.~Br\"autigam, O.E. Josephs, V.~Jousm\"aki, and
  C.D.Tesche.
\newblock Neural processing of human faces: a magnetoencephalographic {MEG}
  study.
\newblock {\em Expt. Brain Res.}, 118:501--510, 1998.

\bibitem{e:Lu91}
S.T. Lu, M.S. H\"am\"al\"ainen, R.~Hari, R.J. Ilmoniemi, O.V. Lounasmaa,
  M.~Sams, and V.~Vilkman.
\newblock Seeing faces activates three separate areas outside the occipital
  visual cortex in man.
\newblock {\em Neuroscience}, 43 (2/3):287--290, 1991.

\bibitem{e:Halgren95}
E.~Halgren, T.~Raij, K.~Marinkovic, V.~Jousmaki, and R.~Hari.
\newblock Magnetic fields evoked by faces in the human brain: 1. topography and
  equivalent dipole locations.
\newblock {\em Society for Neuroscience}, 21:562, 1995.

\bibitem{e:Janday89}
B.S. Janday and S.J. Swithenby.
\newblock The use of the symmetric sphere model in magnetoencephalographic
  analysis.
\newblock In S.N. Erne and G.L. Romani, editors, {\em Advances in Biomagnetism
  Functional Localization: A challenge for Biomagnetism}, pages 153--160,
  London, 1989. World Scientific.

\end{thebibliography}

\end{document}